\documentstyle[11pt,newpasp,twoside,psfig]{article}
\reversemarginpar

\begin{document}
\title{YSO disk structure and planetary signatures}
\author{Nuria Calvet}
\affil{Harvard-Smithsonian Center for Astrophysics, 60 Garden St.,
MS 42, Cambridge, MA -2138}
\author{Paola D'Alessio}
\affil{Instituto de Astronom{\'\i}a, UNAM, Ap.P. 70-264,
04510 M\'exico D.F.}

\begin{abstract}

Detailed radial/vertical structure modeling of observations of disks
around Young Stellar Objects (YSOs) can provide information on the
physical conditions and on the characteristics of the gas and dust in
their interiors.  We describe recent results of self-consistent
modeling of spectral energy distributions, optical and infrared
images, and millimeter fluxes of YSOs.  We discuss observations and
interpretations of the initial stages of planet formation, including
indications of dust growth and settling in disks around young stars.
We also discuss how the unprecedented resolution and sensitivity of
ALMA may help us study the interior of the innermost disks, a region
unaccessible with present day instrumentation, and witness the very
first stages of planet formation.

\end{abstract}

\section{Introduction}

Since the late 1970s, it was known that 
the low mass pre-main sequence  T Tauri stars were surrounded
by disks. For instance,
the lack of correlation between the (near infrared) excess and the
extinction had led to the suggestion that the cool material producing the excess
was located in a geometrically flat configuration
(Cohen \& Kuhi 1979).
However, it was not until the first {\it IRAS} observations of TTS appeared
that the large mid and far infrared excesses were attributed to 
accretion disks (Rucinski 1985), in agreement with the original
suggestion of Lynden-Bell \& Pringle (1974). Masses and radii for these
disks were estimated from submillimeter and millimeter observations
around the late 1980's. The indirect inference for disks from all
these observations was dramatically confirmed in this 
decade, when disks around young stars have been directly
imaged by instruments on board of {\it HTS}. Disks are seen
against the bright background of the Orion nebula
as silhouettes (McCaughrean \& O'Dell 1996; McCaughrean et al. 1998;
Stauffer et al. 1999), and in the favorable edge-on situations,
where the direct light from the star is blocked by the
disks, so the faint glow of stellar light scattered by dust
grains on the disk surface can be seen.  
A number of cases have been presented so far;
the best known and analyzed include HH30 (Burrows et al. 1996) 
and HK Tau/c
(Stapelfeldt et al 1998). 
These disks are the predecessor of planetary systems.

T Tauri disks are composed of gas and dust. The mass in the
disk is of the order of a few 0.01 $M_{\sun}$, and matter is
still
accreting onto the star. On the other extreme, in disks with planetary
systems,  most of the mass is on the planets, a few 0.001 $M_{\sun}$,
and only small amount of dust ($<< 10^{-6} M_{\sun}$) 
remains.
The transition between these
two stages is of upmost interest. What are the 
processes that lead to planet formation and in which timescales?

In this contribution, we argue that the
first signs of the transition to planetary disks
appear in the inner disk, within a few AU from the
star, and then when this
transition starts, it happens very fast. We will
review evidences provided by two groups of
different ages, 1-2 Myr and 10 Myr. We then will
argue that present and future observations in
the optical and infrared cannot tell us what
is happening near the midplane of the inner regions of the gas and
dust disks characterizing the initial stages before
the transition. Only ALMA, in the highest resolution
configuration, will allow us to scrutinize the  innermost
disk and to obtain information on the processes
occurring to the dust in these regions leading
to the formation of planets.

\begin{figure}[ht]
\plotfiddle{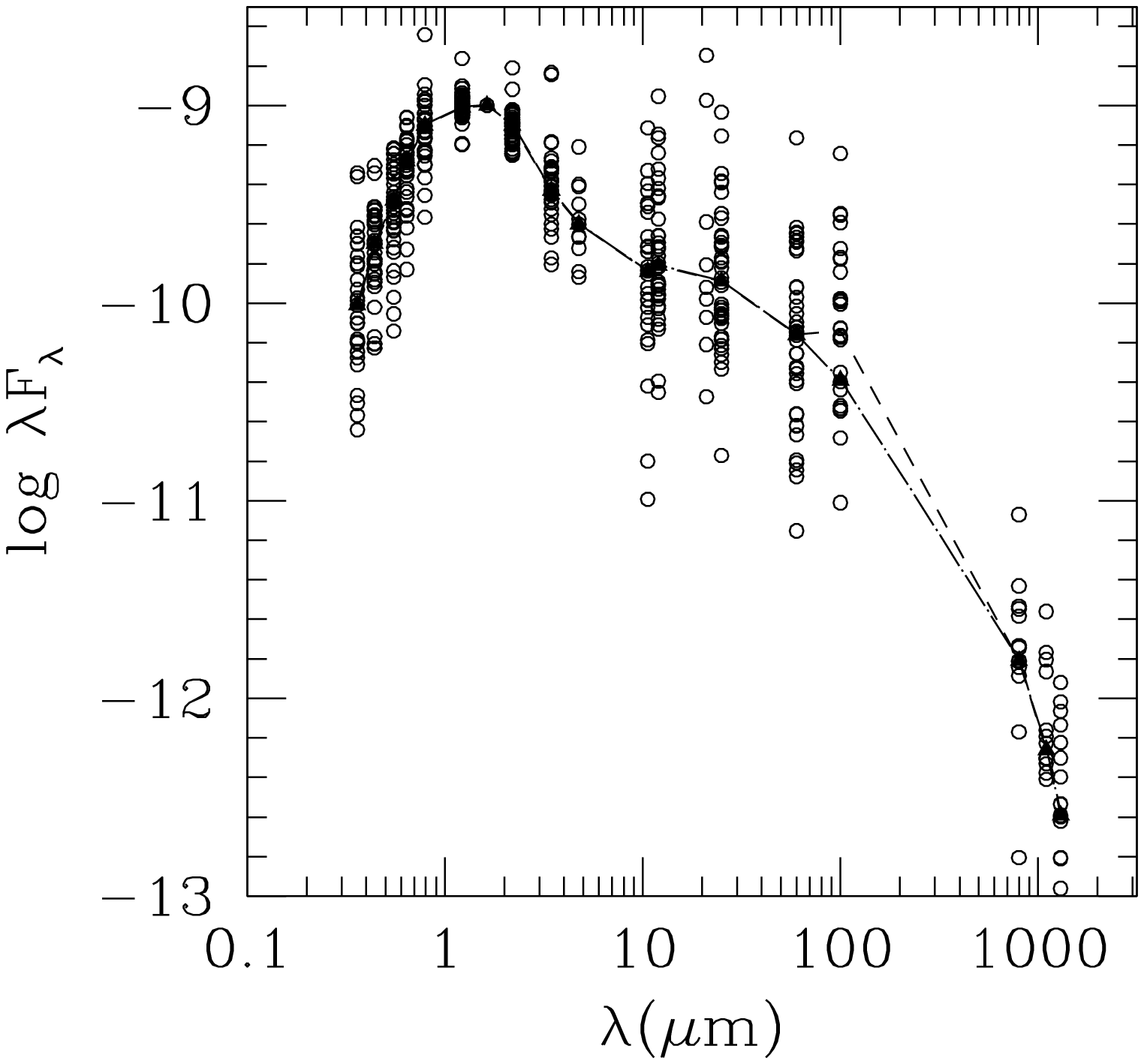}{2.5in}{0}{50.0}{40.0}{-150.0}{-70.0}
\end{figure}

\noindent
Fig. 1. Fluxes of classical T Tauri stars in Taurus-Auriga.
From D'Alessio et al. (1999b.) Fluxes are normalized at H.
The solid line is the mean SED constructed from these data,
including upper limits at 100$\mu$m. The dashed line does
not include these upper limits.

\section{Indications of dust evolution}

Figure 1 shows fluxes for all classical T Tauri stars (CTTS)
in Taurus-Auriga, scaled at H (D'Alessio et al. 1999b).
The large infrared excesses due to disk emission
are apparent.
Disk emission depends on the temperature of the disk, which
in turn, is determined by the amount of heating. Two agents
can heat the disk: viscous dissipation related to accretion, and
stellar irradiation. For the typical values of the mass accretion
rates in the disks around CTTS (Gullbring et al. 1997; Hartmann et al.
1998), irradiation from the central star plays a
dominant role in the heating of the disk (Kenyon
\& Hartmann 1987, KH87;
D'Alessio et al. 1998, 1999a,b,c). 
Heating by irradiation
will depend on the shape of the disk surface, since
disks with flared surfaces can capture more
stellar radiation
than flat disks (KH87). In turn,
the scale height of the disk, and therefore
the height of the surface, depends on the disk temperature.
A self-consistent stability analysis of irradiated
disks for typical CTTS parameters indicates that 
a unique, stable solution for the surface exists,
and corresponds to a flared surface
(D'Alessio et al. 1999a).
This means that 
disk emission will be determined essentially by
how much energy comes into the disk, set by
the star, and how much of this energy is actually
absorbed and re-emitted by the disk, which depends
on the dust in the disk.

Most stars in Taurus
shown in Figure 1 have spectral types between K7 and M3,
so they have similar amounts of stellar irradiation. 
The large dispersion observed in the SEDs must
essentially be due then to differences in the dust properties
in the disk, indicating that even though the stars
have similar ages (1 - 2 Myr, Gomez et al. 1993), 
there is a
large diversity in the conditions of the dust in these disks.

Figure 2 shows the SEDs for two ``transition
cases'' in Taurus: DI Tau (Meyer et al. 1997) and
V819 Tau (M. Meyer, personal communication). These
stars are weak TTS, that is, they are not accreting
mass onto the star. The indications of disk accretion,
i.e., broad emission line profiles, UV excesses, 
which characterize CTTS are
not present in these stars, nor are near-infrared
excesses. However, mid-infrared and even millimeter
flux excesses persists, indicating that remnants of the disk
are still there. Some other transition cases may exist 
and remain to be
discovered, but not in large numbers; the total 
number of transition cases may be less than 10, which when compared
to the total number of TTS in Taurus, $\sim$ 100,
indicates that the loss of the disk happens fast.
The actual processes causing disk dissipation
are not well known, and
binary companions may play a crucial role in
clearing the disks; but in any event, the processes
leading to disk dissipation occur in very short
timescales, compared to the lifetime of a typical T Tauri star.

\begin{figure}[ht] 
\vskip -1in
\plotfiddle{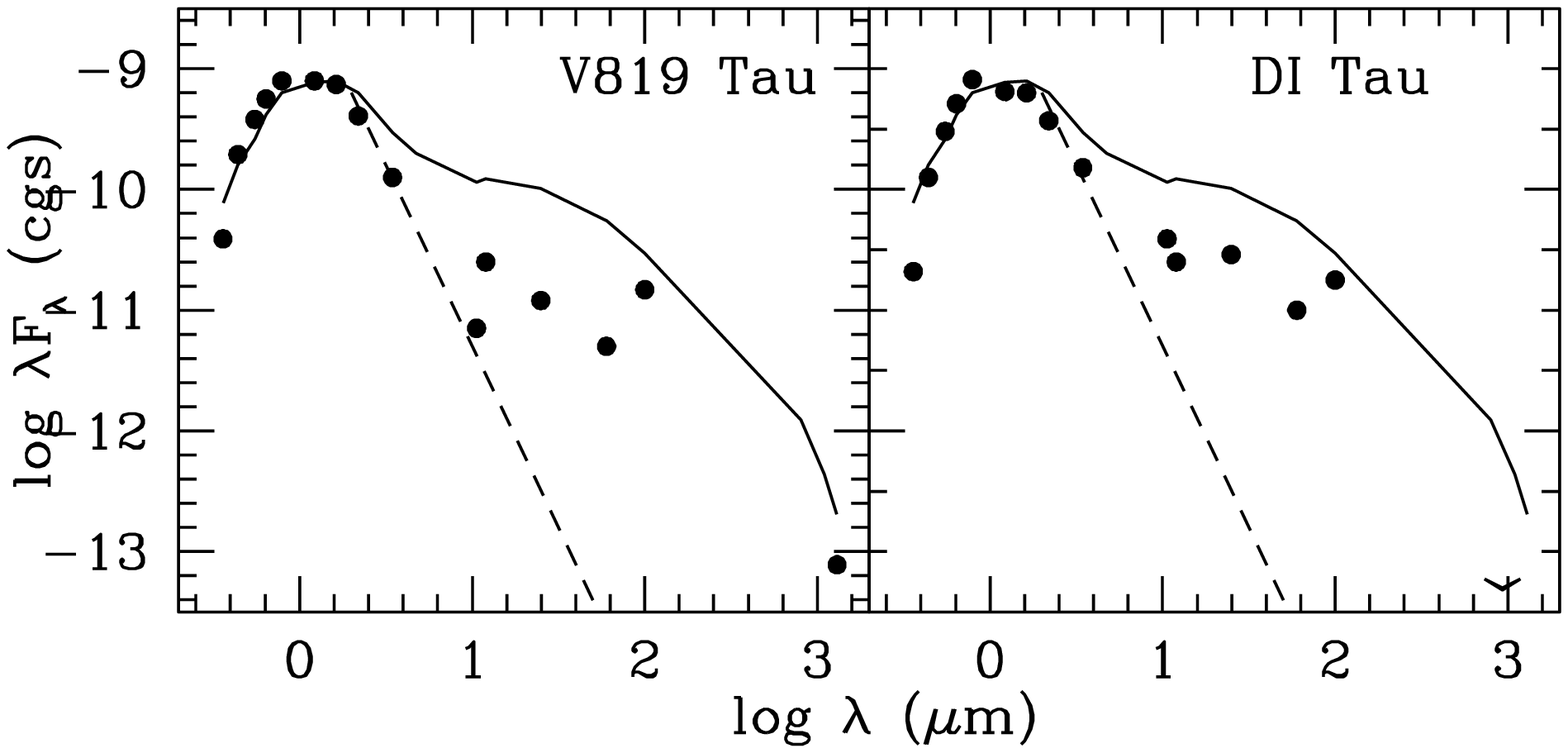}{2.5in}{0}{50.0}{40.0}{-150.0}{-70.0} 
\end{figure} 
 
\noindent
Fig. 2. SEDs of two ``transition'' T Tauri stars in Taurus:
DI Tau (right) and V819 Tau (left) (data from Kenyon
\& Hartmann 1995). These are
weak T Tauri stars, where accretion has already stopped,
but still with mid and far infrared excesses.
The median SED in Taurus (solid line) and
photospheric fluxes (dashed line) are shown for comparison.

\vskip 0.3in

Figure 3 shows SEDs for stars in the TW Hya 
association, which has an estimated age of
10 Myr (Webb et al. 1999) from Jayawardhana et al. (1999).
Although TW Hya itself has been known for many
years (Herbig 1978; Rucinski \& Krautter 1983;
de la Reza et al. 1989) as an ``isolated T Tauri star'', it
is not until recently that this extremely interesting
association has begun to be extensively studied.
The main problem has been finding its members, since
molecular gas is no longer associated with the
association.
Most of the members found are turning out to be
weak TTS, but a few of them still show indications
of disk emission, similar to the mean SED
in Taurus in the mid and
far-IR and in the millimeter (Figure 3). Again, in some cases,
a deficit in the near-IR compared to longer wavelengths
seem to be present.

The evidence provided by young
stars in the two extremes of the temporal range
where disks are expected to dissipate and planets
to form ($\sim$ 10 Myr, Podosek \& Cassen 1994)
indicates that disks can last for a very long
time, but once clearing starts, it occurs in very short
time-scales. The first signs of disk clearing 
seem to appear in the near-IR, indicating that
the higher temperature regions in the inner disk
are the first to clear up. This is is consistent
with the evidence provided by debris disks; for
instance, in the youngest debris disk known,
around the A0 star
HR 4796A, $\sim$ 8 Myr (Jayawardhana et al. 1998;
Koerner et al. 1998), a possible member of the TW Hya association,
disk emission is only apparent beyond 10 microns.
The inner disk has been cleared out maybe
by a planet or planets, consistent with
the images in the mid-IR and in the near-IR from
{\it HST} (Telesco et al. 1999; Schneider et al. 1999). 

\vskip 0.3in 

\begin{figure}[ht]   
\plotfiddle{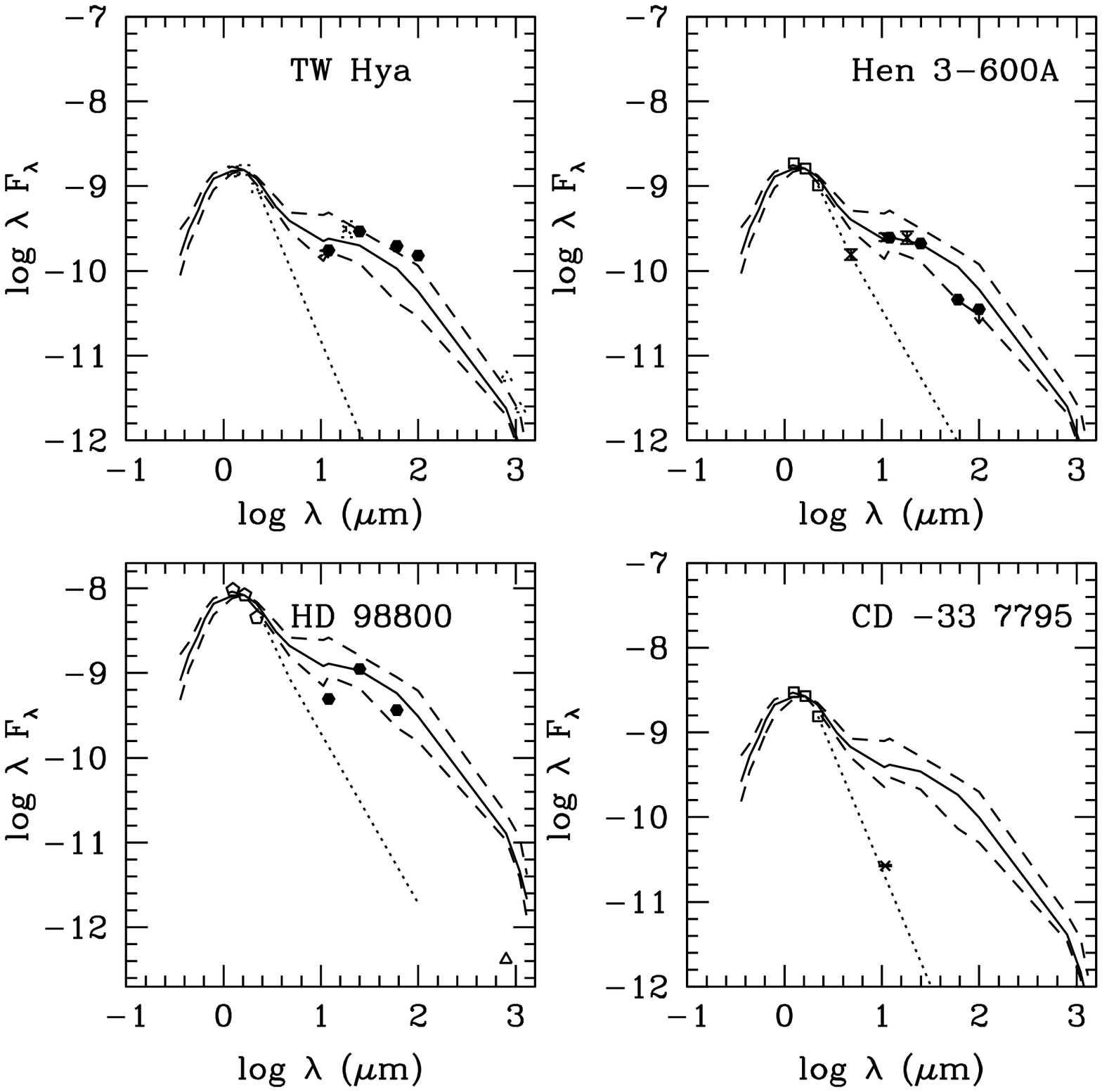}{2.5in}{0}{50.0}{40.0}{-150.0}{-70.0}   
\end{figure}     
  
\noindent 
Fig. 3. SEDs of stars in the TW Hya association.
From Jayawardhana et al. (1999).
The solid curve is the median SED in Taurus,
and the dashed lines indicate the quartiles.
The dotted line indicates the photosphere.

\vskip 0.3in

The process of inner disk clearing must start
with dust growth and settling towards the midplane.
Eventually, a planet (or planets) forms
in the disk, and 
a gap is opened at the position of the 
planet; the inner disk is accreted onto the  
star, while a considerable fraction of the
remaining mass of the disk goes into planet
formation. Disks around CTTS
must be in the very first stages of this
sequence; once planets form   
and gaps appear, the evolution happens very
fast, and it may prove difficult to find 
disks with gaps, still retaining the
inner disk. 
It may bear more results to study the early phases,
previous to planet formation,
when the star still retains
its original gas plus dust disk, 
searching for indication of dust coagulation
and settling.

\vskip 0.3in
 
\begin{figure}[ht]
\plotfiddle{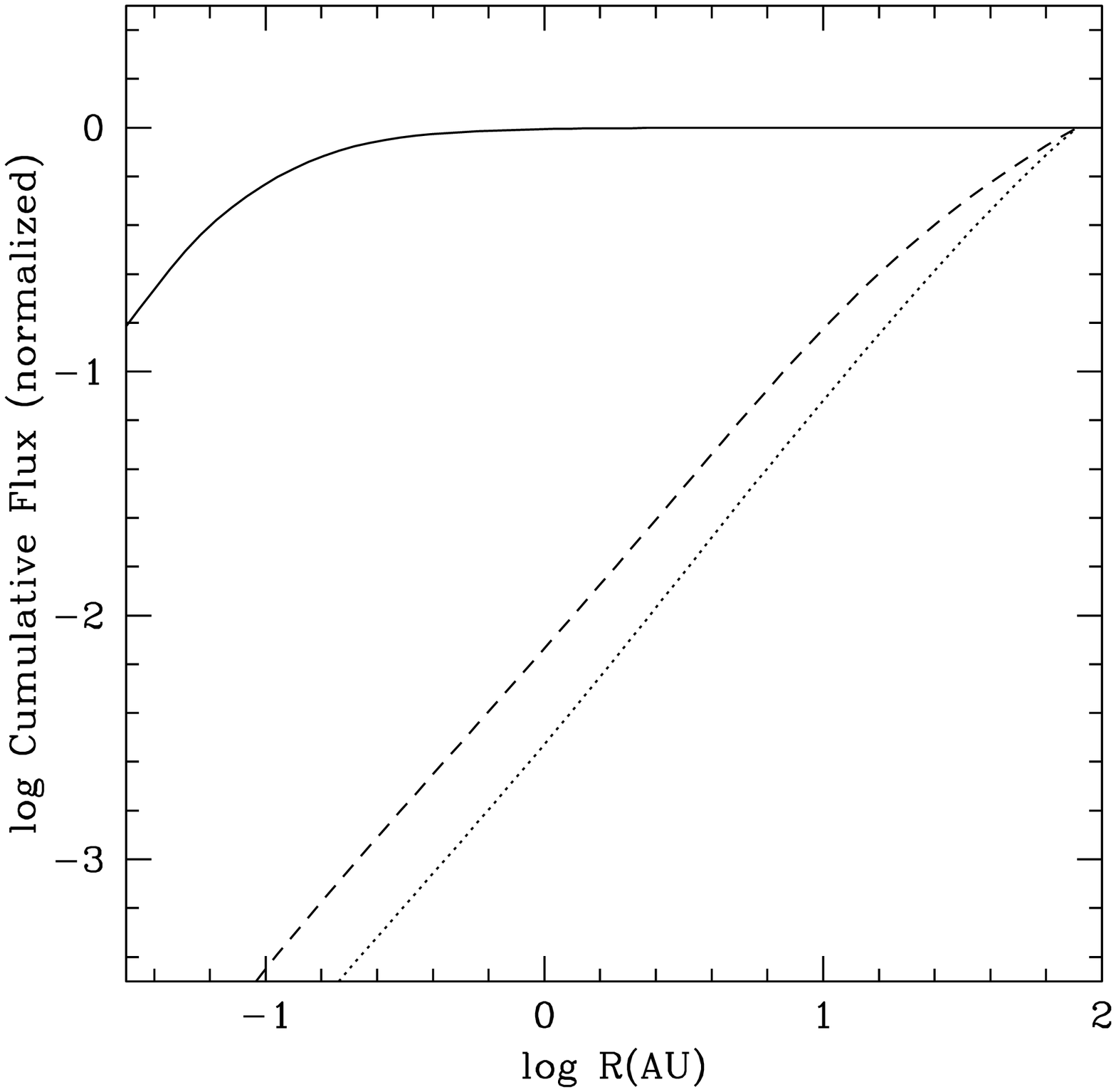}{1.5in}{0}{25.0}{25.0}{-80.0}{-30.0}
\plotfiddle{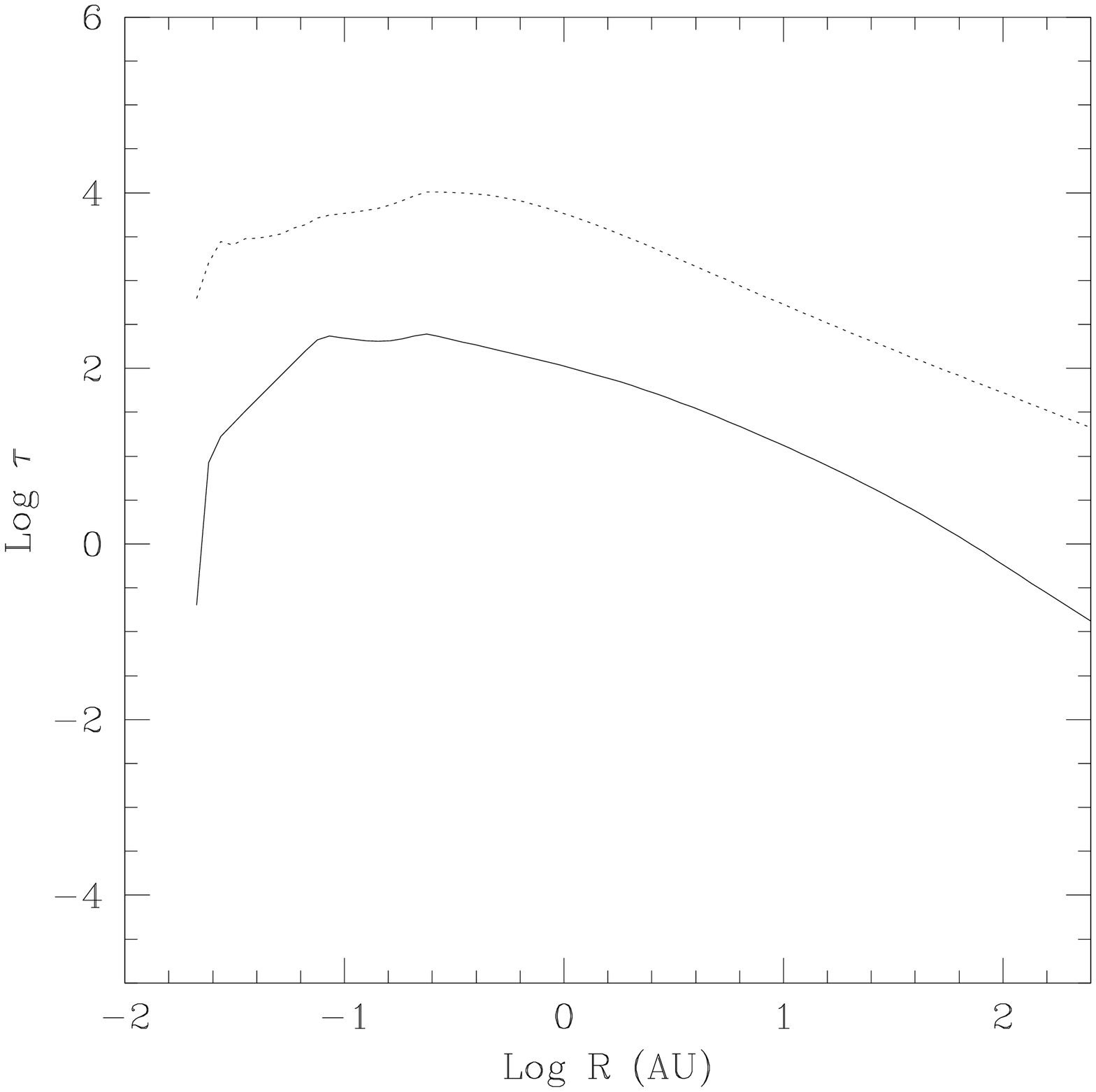}{1.5in}{0}{25.0}{25.0}{-150.0}{-60.0}
\plotfiddle{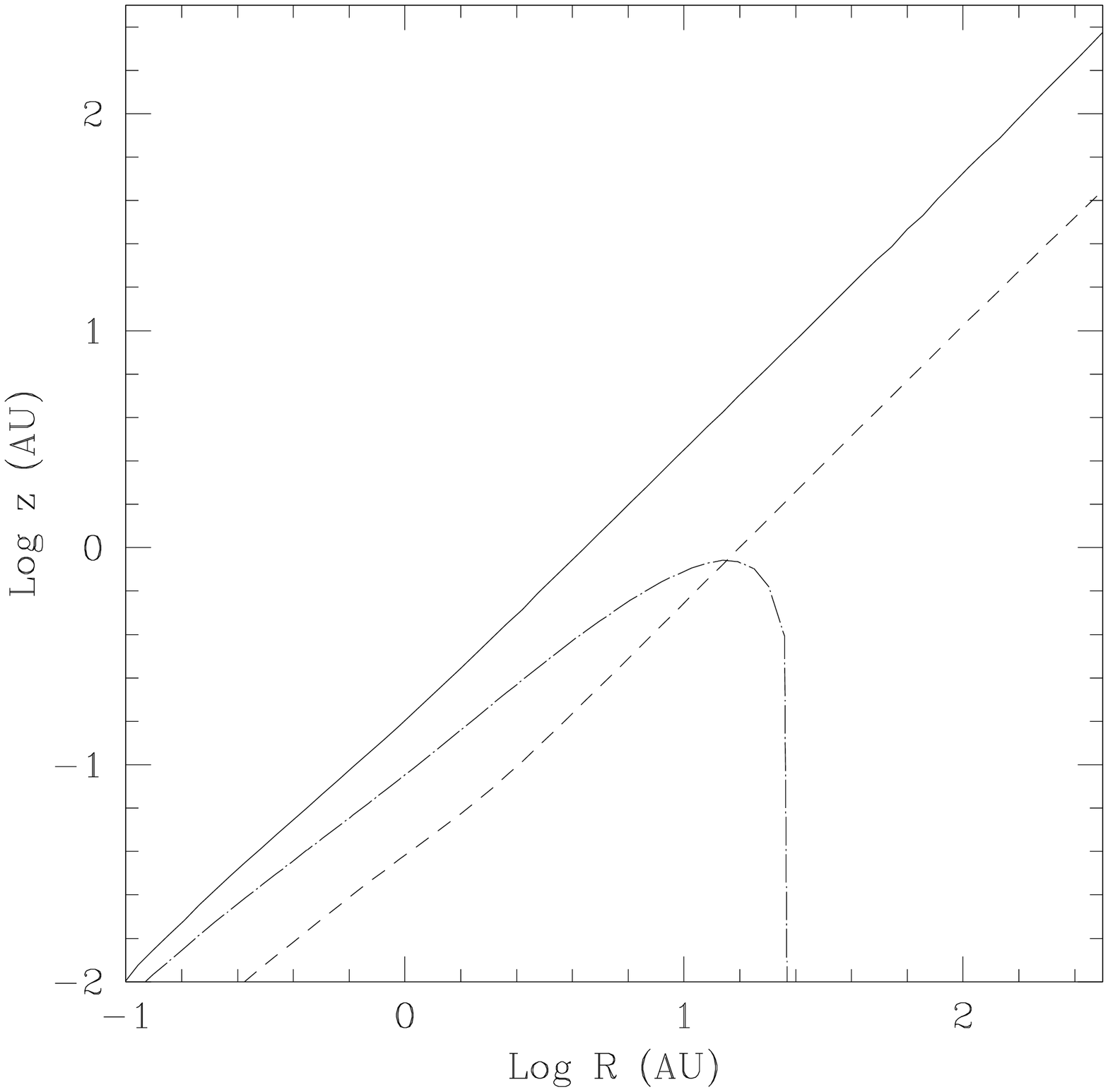}{1.5in}{0}{25.0}{25.0}{-0.0}{60.0}
\end{figure}
 
\vskip -0.9 true in
 
\noindent
Fig. 4. Upper panel: (a) Disk cumulative flux as a
function of radius at three wavelengths:
10 $\mu$m (solid), 0.5 mm (dotted) and 1 mm (dashed).
The disk model has typical
parameters for CTTS: mass 0.5 $M_{\sun}$, radius 2 $R_{\sun}$,
mass accretion rate $10^{-8} M_{\sun} \, yr^{-1}$, and
effective temperature
4000K.
Lower left: (b) Characteristic optical depths in the disk:
$\tau(1\mu$m), where stellar radiation is absorbed (dotted);
$\tau_{Ross}$, the optical depth at the local radiation (solid).
The disk is optically thick to the stellar radiation
even in the outer regions where it becomes optically
thin (to its own radiation).
Lower right: (c) Characteristic heights in the disk:
$z_s$, surface where $\tau(1\mu$m) $\sim$ 1 (solid);
the disk scale height (dashed);
$z_{phot}$, surface where $\tau_{Ross} \sim$ 1
(defined where the disk is optically thick)(long dashes).

\section{The Tauri disks}

Self-consistent radial/vertical calculations of disk
structure for typical CTTS parameters, including viscous
and irradiation heating, have been carried out by D'Alessio
et al. (1998, 1999b,c). These calculations show that
the flux at $ \lambda \le$ 10 $\mu$m, where the first
indications of disk clearing appear, comes from the region
inside a few AU of the disk. Figure 4a shows the 
cumulative flux at three wavelengths, normalized 
at the maximum, indicating that only in the millimeter
range the whole disk is contributing to the flux.

\begin{figure}[ht]
\plotfiddle{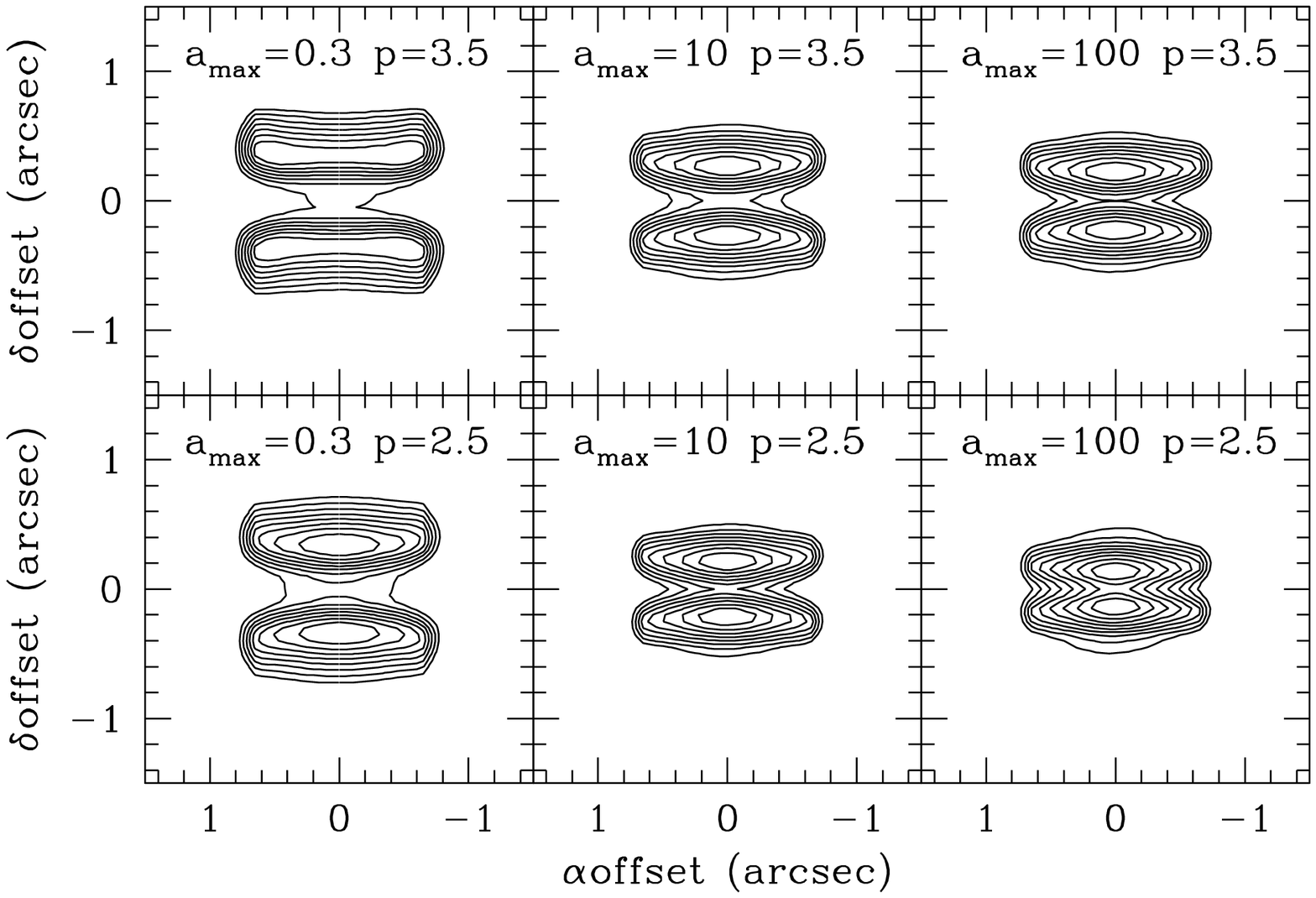}{3.0in}{0}{35.0}{43.0}{-180.0}{-110.0}
\plotfiddle{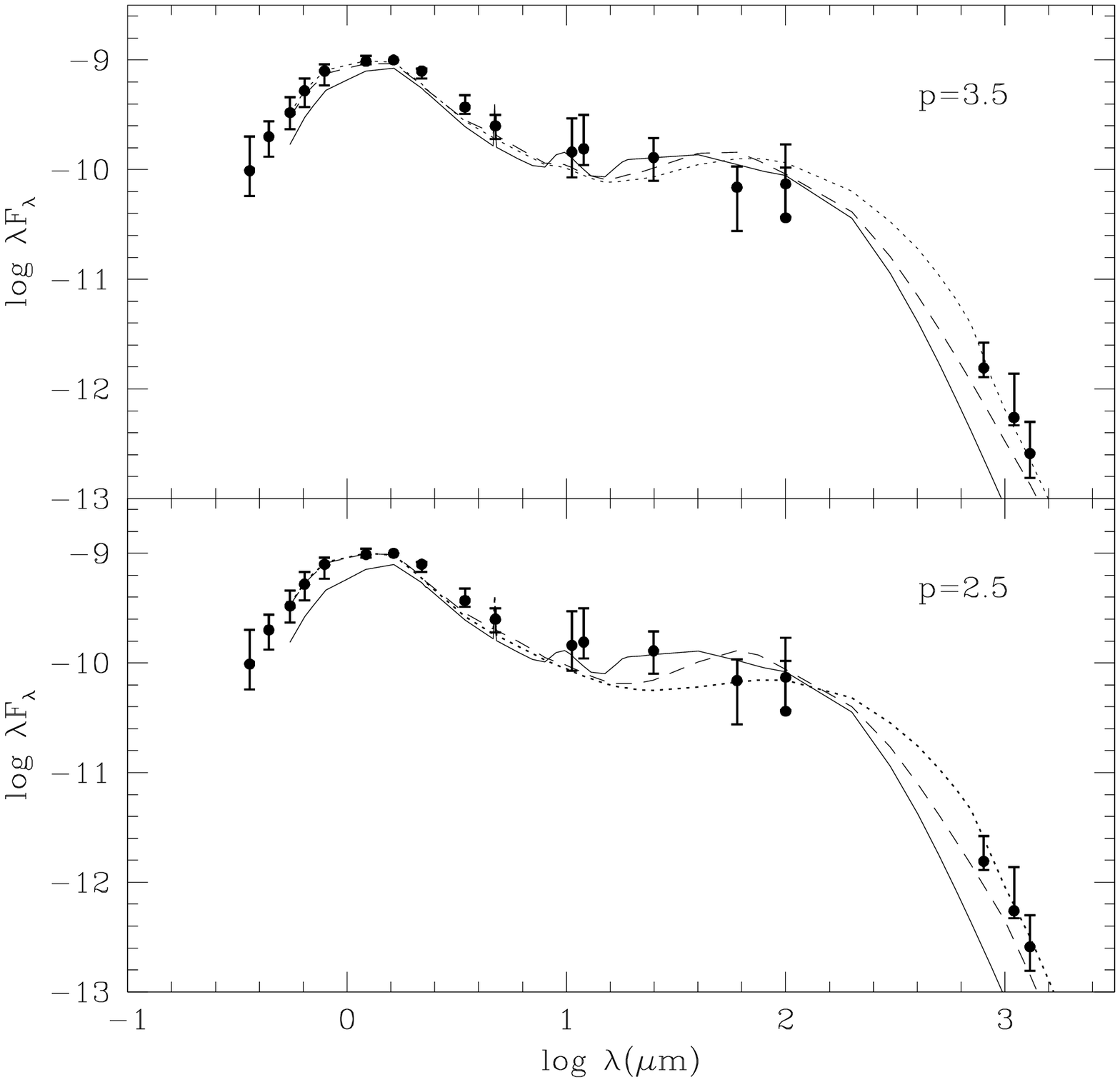}{1.5in}{0}{30.0}{28.0}{20.0}{111.0}
\end{figure}
 
\vskip -1.8in
\noindent
Fig. 5. Left: {\it HST} scattered light model images
at 0.8$\mu$m for edge-on
disks, corresponding to dust well mixed with gas and
different size distribution and maximum size (see text).
Right: SEDs for models with images in the left size.
$a_{max} = 0.3 \mu$m (solid), 10 $\mu$m (dashed), and 100 $\mu$m
(dotted).  From D'Alessio et al. (1999c).

\vskip 0.3in

Figure 4b shows characteristic optical depths in the
disk. The optical depth at $\lambda \sim 1 \mu$m,
at the peak of the stellar radiation, is always
much larger than 1,
since dust opacity is high at optical and
near-IR wavelengths. In contrast, the Rosseland
mean optical depth, which corresponds
to the optical depth
to the local radiation, falls below 1 beyond $\sim$
10 AU. This means 
that the disk is thick
to stellar radiation even when it is optically
thin to the local radiation (for instance, in the outer disk.)
This last property implies that the disk 
can efficiently capture stellar radiation
up to a few hundred AU, of the order of
the estimated radii of CTTS (Beckwith et al. 1990;
Dutrey et al. 1996). It also implies that
scattered light images of edge-on disks
observed with {\it HST} should show the flared
shape characteristic of the surface where
stellar irradiation is being captured, as
it is indeed the case. Figure 4c shows this
surface, $z_s$, as well as the scale height
and $z_{phot}$, the
height of the optically thick disk (obviously defined 
only in the region where the disk is thick).

\vskip -1.5in

\begin{figure}[ht]
\plotfiddle{ 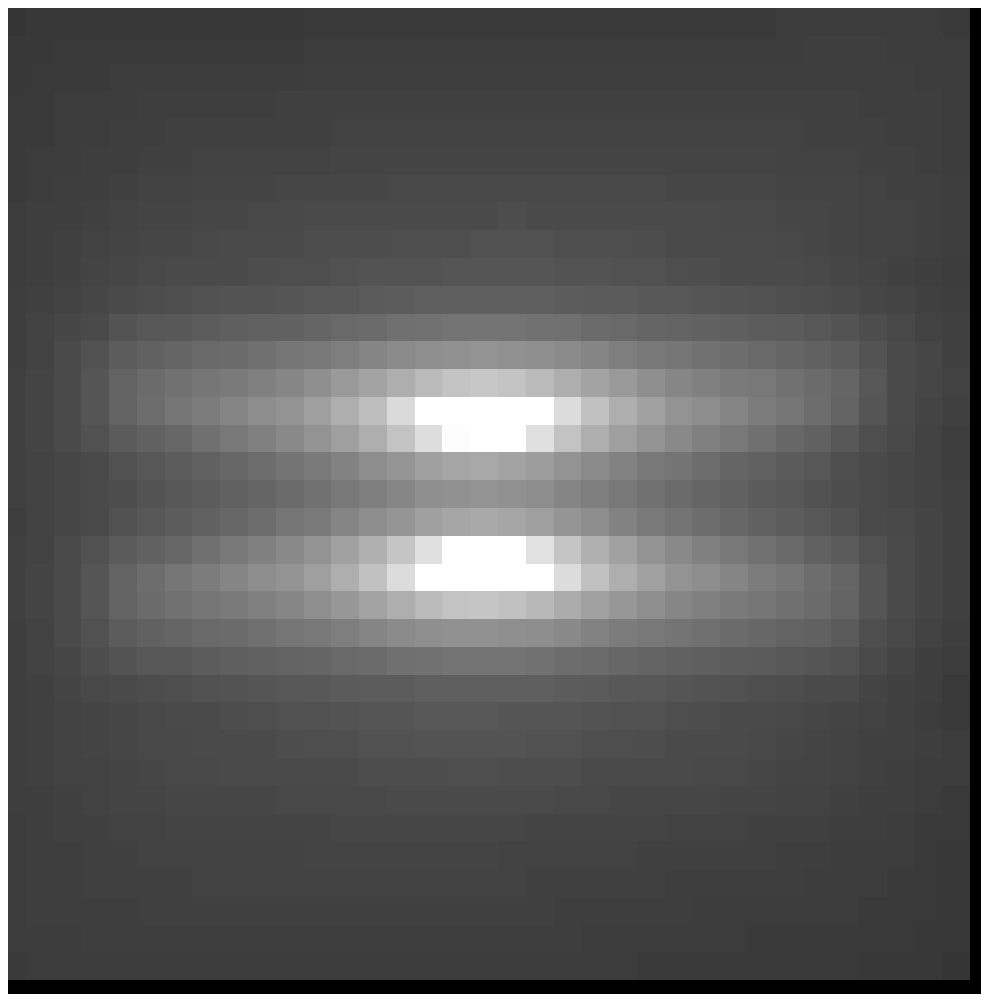}{4.0in}{0}{45.0}{35.0}{-200.0}{-28.0}
\plotfiddle{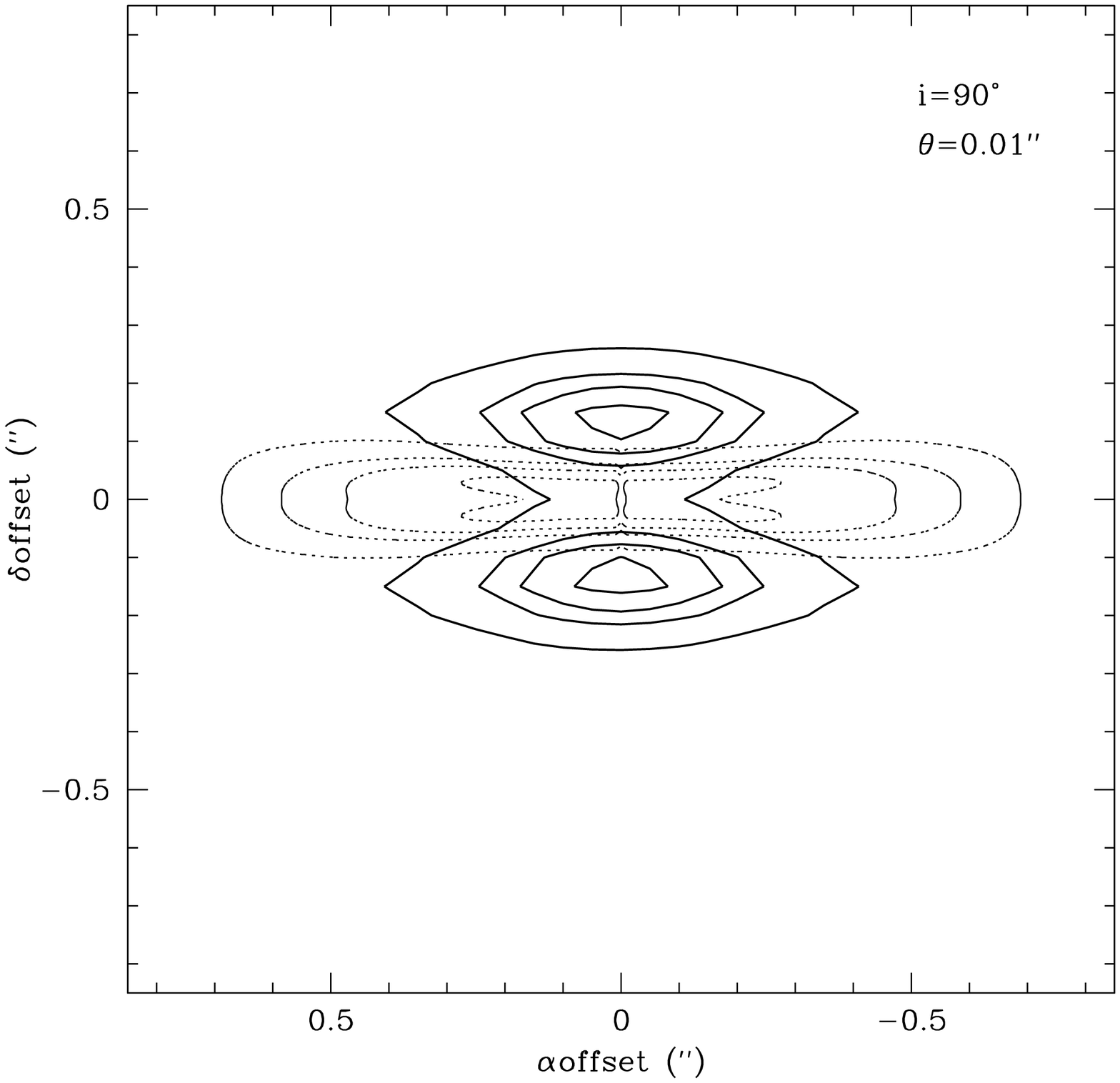}{1.5in}{0}{30.0}{28.0}{20.0}{111.0}
\end{figure}

\vskip -1.9in
\noindent
Fig. 6. Left: Model scattered light {\it HST} image at 0.9 {\AA}
for typical CTTS parameters.
Right: isocontours of image on left (solid), superimposed
to disk emission observed with ALMA
at 0.9 mm with
0.01" resolution (dotted).

\vskip 0.3in

{\it HST} scattered light images and 
SEDs of CTTS may
already be given indication of dust growth
in the disk. Figure 5 shows calculations
of images and SEDs for
a mixture of astronomical silicates
and graphite (Draine \& Lee 1987, DL87), with
a size distribution $n(a) \propto a^{-p}$,
with $p =$ 3.5 (MRN, typical of interstellar medium dust,
DL 87) and $p =$ 2.5 (indicating some degree of
coagulation, Miyake \& Nakagawa 1995). The
minimum size of the dust is $a_{min} = 0.005 \mu$m (DL87)
and the maximum size is $a_{max}$ = 0.3 $\mu$m (DL87, ISM
dust),
10 $\mu$m,  and
100 $\mu$m, for fixed dust to gas ratio. As the
maximum grain size increases, the opacity decreases in the
optical and near-IR, as more of the dust is locked
in bigger grains; at the same time, the opacity
increases in the millimeter range. The effect
of grain growth on scattered light images is
shown in Figure 5, left. ISM dust produces images 
where the widths of the
lanes separating the bright nebulae above and below the
disk are of the order of $\sim$ 100 AU (at Taurus), 
which is larger than the width of the
observed edge-on disks (D'Alessio et al. 1999b). 
The width of the lane decreases as $a_{max}$
increases, as a result of the decreased opacity
in the optical and the consequent lowering of the
surface where $\tau (1 \mu$m) $\sim 1$, producing
better matches to the observations.
At the same time,
mixtures with larger grains produce much better
agreement between the typical CTTS disk model
fluxes and
the median Taurus SED (Figure 5, right).

\section{Scrutiny of the deep inner disk: only with ALMA}

{\it HST} observations, even at $\sim$ 0.05" resolution,
can only probe the upper regions of the disk, because
the CTTS disk is {\it always optically thick to optical and
near-IR radiation}. Only submillimeter and
millimeter observations can reach the mid-plane of the
disk. But present day interferometers can
only give us a resolution of at most $\sim$ 40 AU at the
closest star formation regions. This implies
that we have no direct 
observations to constraint the detailed structure of the 
regions near the midplane in the inner disk
regions. ALMA at the highest
resolution, $\sim$ 0.01", will constitute the
perfect complement to {\it HST} observations, allowing us
to reach the true scale height of the disk.
To illustrate this,
Figure 6 compares typical CTTS model calculations for
a scattered light {\it HST} edge-on disk image at 0.9 {\AA} and
the corresponding ALMA image at 0.9 mm with
0.01" resolution.

Moreover, with the highest ALMA resolution, 0.01" $\sim$ 1.4 AU
at Taurus, we will be able to probe different
heights at the innermost disk regions observing
at different wavelengths. Figure 7, left, shows
the heights in the disk where $\tau \sim 1$
at different wavelengths. In the submillimeter
range and beyond, the disk opacity is low enough
to allow inspection of regions closer and closer
to the midplane. These observations will allow
us for the first time to {\it effectively map the inner disk
structure}.

\vskip -0.8in

\begin{figure}[ht]
\plotfiddle{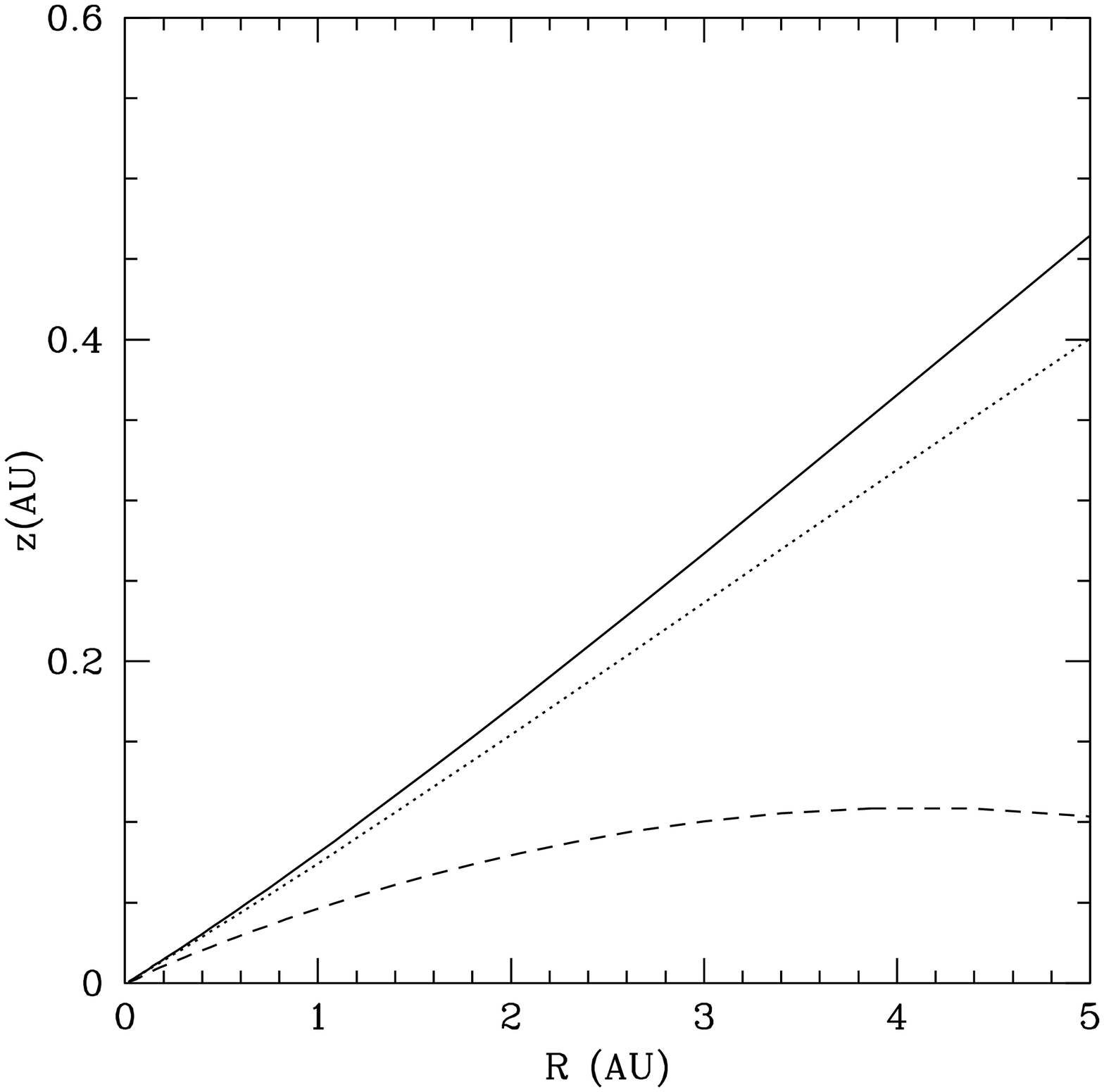}{2.5in}{0}{30.0}{30.0}{-180.0}{-80.0}
\plotfiddle{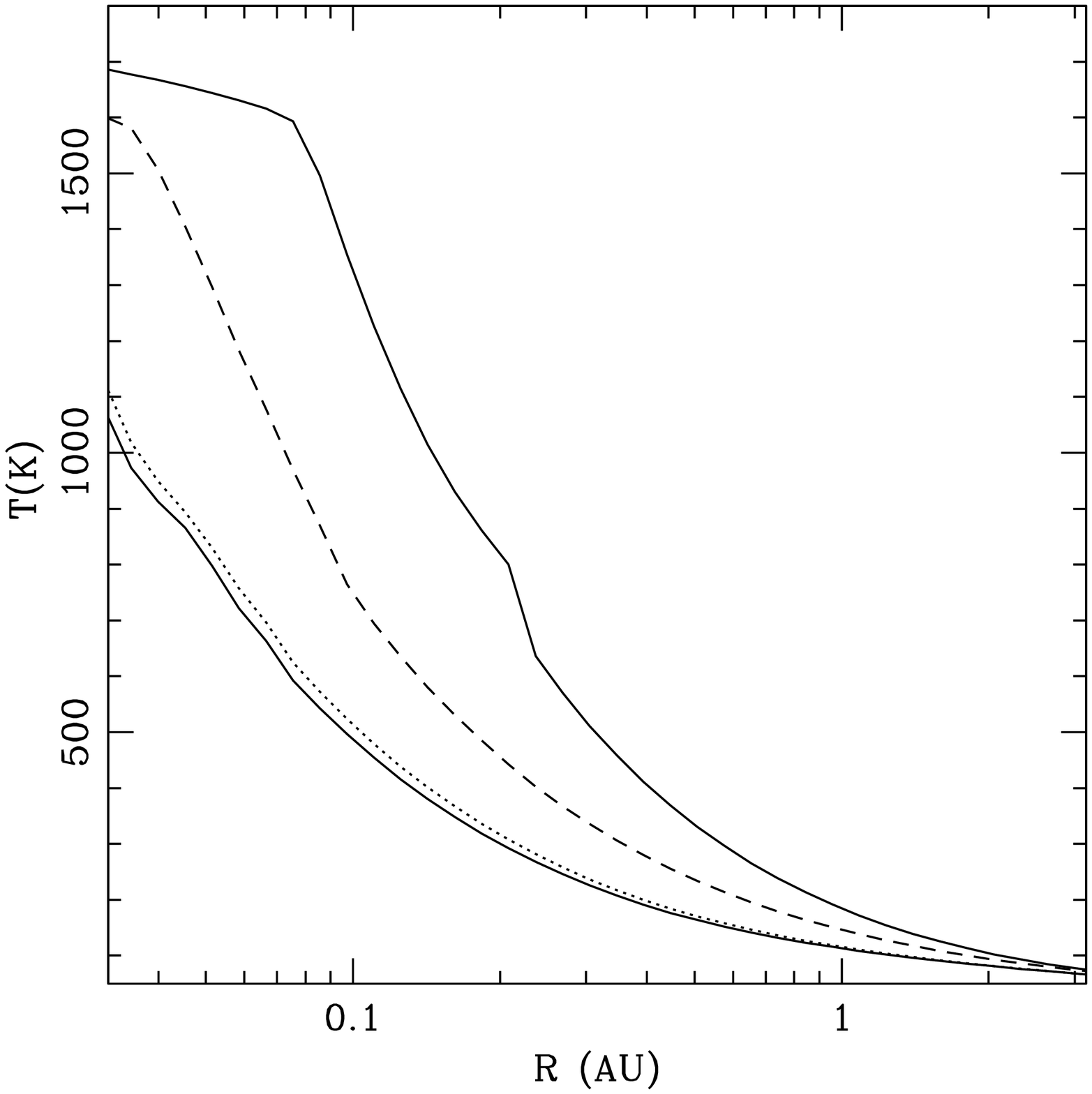}{2.5in}{0}{30.0}{30.0}{20.0}{111.0}
\end{figure}

\vskip -1.9in
\noindent
Fig. 7. Left: Height in AU at which the optical depth
becomes 1 at given $\lambda$, i.e., height of formation
of radiation at $\lambda$, for the case
of gas and dust well mixed and $a_{max} = 100 \mu$m.
Light solid: 0.5$\mu$m,
dotted: 0.5 mm, dashed: 1 mm.
Right: Brightness temperature at different $\lambda$,
corresponding to heights in left. The heavy solid
line is the midplane temperature.

\vskip 0.3in
 
The temperature in the inner disk, $T(z,R)$
depends on local conditions. Using the diffusion
approximation, $dT/dz \propto \kappa F$. The
opacity $\kappa$ will depend on the dust
properties, type, size, and vertical distribution.
The flux generated at each height, $F$, will
depend on the local viscous generation
of energy. Figure 7, right,
shows the brightness temperature 
at several wavelengths,
for the case
of dust with $a_{max} = 100 \mu$m, well mixed
with gas, and viscous heating at each height. 
Because of the large opacity,
the temperature in the mid-plane is much
higher than the surface temperature. Brightness temperatures
in the millimeter are consequently much higher than
those obtained if the disk were isothermal.
However, the actual brightness temperatures
may be very different 
if there is a vertically-dependent dust opacity,
due to dust coagulation and settling.
Moreover, they may be very different if
a ``dead zone'' exists near the mid-plane,
where viscous generation of energy is not 
occurring because there is not enough 
ionization to maintain the magneto rotational
instability (MRI, Gammie 1996; Glassdgold et al. 1997).
The importance of this dead zone for planet
formation is that inhibition of the MRI
produces an increase of the 
density near the midplane (since matter cannot
be transported inwards towards the star), 
favoring gravitational instabilities and
potentially leading to planet formation.

\section{Conclusions}

The present evidence from disks around 1 to 10 Myr
old stars indicates that dust growth and
settling leading to planet formation and disk clearing happens fast
and starts near the midplane in the few inner AU of the disk.
Since the inner disk regions are optically thick
to radiation shortwards of $\sim$ 0.5 mm, only ALMA
in the highest resolution configuration, corresponding to
0.01" or 1.4 AU at the closest star-forming regions,
will allow us to reach the midplane of the planet-forming
regions in the disks and study the
processes that precede the formation of these bodies.

\acknowledgements
This work was supported in part by NASA grants NAGW 2306 and NAG5-4282.

\end{document}